\documentclass[a4paper]{article}  
\usepackage{graphicx}              

\begin{document}

\title{The G\"odel Universe: A Practical Travel Guide}
\author{
\\
Deshdeep Sahdev\\
Ravishankar Sundararaman\\
Moninder Singh Modgil\\
\emph{Department of Physics, IIT Kanpur} \\
} 

\date{Sep 28, 2006}
\maketitle 

\begin{abstract}
We study the G\"{o}del universe through worldlines associated with motion at constant speed and constant acceleration orthogonal to the instantaneous velocity (WSAs). We show that these worldlines can be used to access every region --- both spatial and temporal --- of the space-time. We capture the insights they accord in a series of sketches, which extend significantly the Hawking and Ellis picture of the G\"{o}del universe. 
\end{abstract}

\section{Introduction}

The G\"{o}del Universe\cite{GodelOrigPaper} has several remarkable properties. It is homogeneous (manifestly so in all studied versions of `rectangular' coordinates) rotationally symmetric (quite obviously, when viewed from an axisymmetric cylindrical system), anti-Machian (owing to its rotation as a rigid body in an inertial frame) and acausal (because its light-cone structure gives rise to closed null and closed time-like curves (CNCs/CTCs)).

Its geodesics\cite{KundtOrigPaper,GenGeodesicPapers} inevitably reflect one or more of its underlying properties. In both the cylindrical and Kundt's version of rectangular coordinates, they project onto circles in the $(r, \phi)$ and $(x,y)$ planes respectively (their closure being a direct consequence of the universal rotation) and correspond, quite generally, to motion at a constant speed, $v$. As $v$ increases, the radius of the circle increases as well, but remarkably, the time taken to circumnavigate it completely, as measured by an observer situated at any one of its points, decreases!  The largest geodesics are null curves, stretching out to the G\"{o}del horizon which marks the boundary of causal behaviour as also of the region which the origin can access through geodesics. 

By subjecting our projectiles to a uniform acceleration, $A$, orthogonal to their instantaneous velocities, we can force them onto orbits of ever larger radii and can thereby bring them back to the base-station ever earlier. So much so that for a critical acceleration, $A_{CTC}(v)$, we could bring them back instantly, and for an even larger $A$, could bring them back before they were launched!

All of these properties can be depicted diagrammatically and some have in fact already been so captured. In an enlightening picture of the G\"{o}del universe in cylindrical coordinates, Hawking and Ellis\cite{HawkingEllis} brought out beautifully the refocussing of photons diverging
uniformly from an arbitrarily chosen origin, the symmetrical tilting of light-cones in the direction of rotation, the emergence of CNCs
at the G\"{o}del horizon and their turning into CTCs thereafter. Pfarr\cite{Pfarr} took the first step towards developing a similar sketch in Kundt's coordinates, by tracing the corresponding geodesics, CNCs and CTCs along with their planar projections.

In this paper, we take Pfarr's discussion to its logical conclusion. We start by showing that the extensions of geodesics he considered, correspond to motion at uniform speed and magnitude of acceleration, orthogonal to the instantaneous velocity. The WSAs, as we shall refer to these extensions, have all the symmetries of geodesics and in particular the property of refocussing, which now gives rise to a truly intriguing series of diagrams: Indeed, as we approach $A_{CTC}(v)$, the points of divergence and re-convergence draw ever closer, and finally coincide as $A=A_{CTC}(v)$. As we increase $A$ further, we eventually end up with open trajectories, and for even higher $A$, trajectories which curve oppositely to the geodesics. Trajectories for these values of $A$ have not, as yet, been studied in the literature. In fact, WSAs provide a very convenient and insightful way of studying all the interesting phenomena associated with the G\"odel universe.

We also present ways of visualising the general mechanism for the emergence of CNCs/CTCs. Indeed, by explicitly drawing a set of
representative light-cones along the latter, we show how the tilting of these cones along with their opening out to include a spatial axis (in addition to the temporal one) allows us to complete the CNC/CTC. We first show this for cylindrical coordinates and then, in the backdrop of a manifest translational symmetry, for Kundt's coordinates.

Finally, we note that $v$ and $A$ are the control variables so to say for navigating G\"odelian space and time. By supplying the formulae and a diagram to show how these can be adjusted to access a coordinate time of choice, within a specified interval of proper time, we arrive at a practical travel guide for the G\"odel Universe.

The plan of the paper is as follows: In section \ref{sec:cylGeodesic}, we set up the equations of motion for planar geodesics passing through the origin in cylindrical coordinates and show that the geodesics have a constant speed. Further, the geodesic of speed $v$ projects onto a circle in the $R$-$\phi$ plane of diameter $D=v/\sqrt{2}$. In section \ref{sec:WorldlineConstruction}, we obtain a family of worldlines from the geodesics by stipulating that the diameter, $D$, of the trajectory be independent of its speed, $v$. We argue that the worldlines so obtained correspond to motion at constant speed and magnitude of acceleration. We calculate this acceleration by comparing each accelerated worldline with a geodesic which shares the same trajectory in the $R$-$\phi$ plane. In section \ref{sec:cylTIntegral}, we determine the coordinate time, $t$, and proper time, $s$, taken to reach a given point on the trajectory of a worldline. We use these results in section \ref{sec:cylWorldlineProperties} to list several noteworthy properties of these worldlines, which we additionally capture diagramatically in section \ref{sec:cylVisuals}. One of these diagrams lays bare an intuitive mechanism for the emergence of CTCs, which generalizes, in a certain sense, the one captured by Hawking and Ellis. In section \ref{sec:kundtKinematics}, we rederive several of the above results in Kundt's coordinates. We moreover use the manifest homogenity of this coordinate system to remove from the orbits of section \ref{sec:cylGeodesic} the constraint that they must pass through the origin. In section \ref{sec:kundtTIntegral}, we mimic the results of section \ref{sec:cylTIntegral} in Kundt's coordinates. In section \ref{sec:kundtVisuals}, we visualize these results from several perspectives. Finally, in section \ref{sec:summary}, we summarize our findings.

\section{Kinematics - Cylindrical Coordinates} \label{sec:cylGeodesic} 

Our first objective is to construct `planar' wordlines in the G\"odel universe (i.e. worldlines on the plane orthogonal to the $z$-axis) which start from the origin and travel with a constant speed and magnitude of acceleration. Such worldlines are considered because they fully exploit the symmetry of the universe, they are reasonably simple to handle and they are sufficient to illustrate the strange properties of the G\"odel universe. 

We first construct these worldlines in the cylindrical coordinate system $(t,r,\phi)$ \cite{GodelOrigPaper}, for which the line element is:
\begin{equation}
ds^2 = \left(dt+2\sinh^2\tilde{r} \frac{d\phi}{\omega}\right)^2 
- \left( dr^2 + \sinh^2\tilde{r} \cosh^2\tilde{r} \frac{2 d\phi^2}{\omega^2} \right) 
\end{equation}
where $\tilde{r}=\omega r/\sqrt{2}$ and $\omega$ is the angular velocity of rotation of the universe. The $dz^2$ term has been dropped as we consider only trajectories restricted to the plane of rotation of the universe (the $r$-$\phi$ plane). The calculation simplifies considerably 
\cite{OzSchuGodelTrip}, if we rewrite the metric in terms of $R=\tanh(\tilde{r})$ (which maps the entire space-time onto the interior of the unit circle):
\begin{equation}
ds^2 = \left(dt+ \frac{2R^2 d\phi}{\omega(1-R^2)} \right)^2 - \frac{2\left( dR^2 + R^2 d\phi^2 \right)}{\omega^2(1-R^2)^2}
\end{equation}

The effective Lagrangian for a particle moving geodesically is then 
\begin{equation} \label{eqn:CylLagrangian}
\mathcal{L} = g_{\mu\nu}\dot{x}^{\mu}\dot{x}^{\nu}
= \left(\dot{t}+ \frac{2R^2 \dot{\phi}}{\omega (1-R^2)} \right)^2 - \frac{2\left( \dot{R}^2 + R^2 \dot{\phi}^2 \right)}{\omega^2(1-R^2)^2} = \epsilon
\end{equation}
where $\epsilon = 0/1$ for photons/material particles and $\dot{x}=dx/d\lambda$. For material particles, $\lambda = s$.

Since $\mathcal{L}$ is independent of $t$ and $\phi$, there are two constants of motion: 
\begin{equation} \label{eqn:constantpt}
p_t = \frac{1}{2}\left(\frac{\partial\mathcal{L}}{\partial\dot{t}}\right) = \dot{t}+ \frac{2R^2}{1-R^2} \frac{\dot{\phi}}{\omega} 
\end{equation}
\begin{equation} \label{eqn:constantpphi}
p_{\phi} = \frac{\omega}{2}\left(\frac{\partial\mathcal{L}}{\partial\dot{\phi}}\right) = \frac{2R^2}{1-R^2} \left(\dot{t} - \frac{1-2R^2}{1-R^2} \frac{\dot{\phi}}{\omega} \right)
\end{equation}
Note that $p_{\phi}=0$ for geodesics passing through the origin, while $p_t$ is related to the speed of the geodesic. Indeed, the speed of a moving particle as seen by comoving observers is given by \cite{LandauLifshitz}
\begin{equation} \label{eqn:speeddefinition}
v^2 = \frac{(g_{0i}dx^i)^2-g_{00}g_{ij}dx^idx^j}{(g_{00}dx^0+g_{0i}dx^i)^2}
= 1 - g_{00} \frac{g_{\mu\nu}\dot{x}^\mu \dot{x}^\nu}{(g_{00}\dot{x}^0+g_{0i}\dot{x}^i)^2}
\end{equation}
where $i$, $j$ index only the space coordinates. In the present case, using $g_{00}=1$, (\ref{eqn:CylLagrangian})
and (\ref{eqn:constantpt}), we get a constant
\begin{equation} \label{eqn:GeodesicVelocity}
v^2 = 1 - \frac{\mathcal{L}}{\left( \dot{t}+ \frac{2R^2\dot{\phi}}{\omega (1-R^2)} \right)^2} = 1 - \frac{\epsilon}{p_t^2}
\end{equation}

Eliminating $\dot{t}$  between (\ref{eqn:constantpt}) and (\ref{eqn:constantpphi}) to obtain $\dot{\phi}=\omega(1-R^2)p_t$ and substituting this and (\ref{eqn:constantpt}) into (\ref{eqn:CylLagrangian}) to obtain $\dot{R}=\omega(1-R^2)p_t\sqrt{\frac{1}{2}(1-\epsilon/p_t^2)-R^2}$, we get
\begin{equation}
\frac{dR}{d\phi} = \dot{R}/\dot{\phi}
= \sqrt{\frac{v^2}{2}-R^2}
\end{equation}
which readily integrates to
\begin{equation} \label{eqn:geodesicorbit}
R = \frac{v}{\sqrt{2}} \sin (\phi-\phi_0)
\end{equation}

It follows that geodesics which pass through the origin are circles in the $R$-$\phi$ plane with diameter $D=v/\sqrt{2}$. Solution (\ref{eqn:geodesicorbit}) corresponds to a projectile from the origin with initial speed $v$ in the $\phi_0$ direction.  
The maximum radial coordinate it attains is $R_{max}=v/\sqrt{2}$, ie. $r_{max} = (\sqrt{2}/\omega) \tanh^{-1}(v/\sqrt{2})$, which for photons becomes $r^0_{max} = (\sqrt{2}/\omega) \tanh^{-1}(1/\sqrt{2}) = (\sqrt{2}/\omega) \sinh^{-1}1$. No geodesic from the origin can go beyond the circle $r = r^0_{max}$, the G\"odel Horizon. 

\begin{figure}
\begin{center}\parbox{80mm}{\includegraphics[width=80mm]{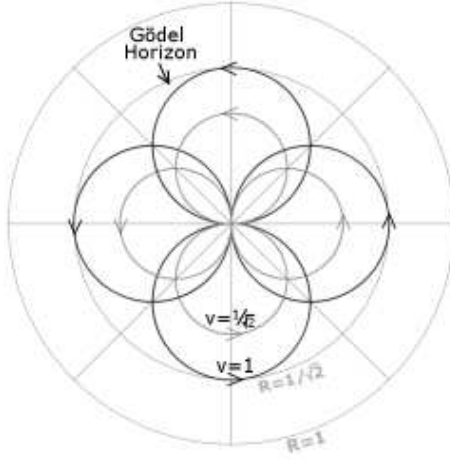}}\end{center}
\caption{The $R\phi$-projections of the geodesics for $v=1$ and $v=1/\sqrt{2}$. The G\"odel Horizon corresponds to $R=1/\sqrt{2}$ and spatial infinity to $R=1$.Note that in these coordinates, this plane is isotropic: particles diverging from the origin with the same speed describe circles of the same radii.  }
\label{fig:cylGeodesics}
\end{figure}

\subsection{Worldlines of Constant $v$ and $A$} \label{sec:WorldlineConstruction}
To see how geodesics can be extended to worldlines with a constant acceleration, it is instructive to examine them in the non-relativistic limit. For $v\ll1$, the geodesic trajectories given by (\ref{eqn:geodesicorbit}) are restricted to $R\ll1$. In this limit, $R\approx\tanh^{-1}R = r\omega/\sqrt{2}$ and hence the trajectories in the $r$-$\phi$ plane are approximately $r = (v/\omega) \sin (\phi-\phi_0)$, that is, circles of diameter $d=v/\omega$. This result can be obtained from a different perspective:

The comoving frame of reference $(\partial_t,\partial_r,\partial_\phi)$ is not inertial. At any given point, it looks like a rotating frame of angular velocity $\omega$, with the axis of rotation passing through that point.\footnote{Every point can appear to be on the axis of rotation only in a curved spacetime. Conventionally, the non-relativistic limit of the G\"odel solution is taken by fixing a rotation axis, and introducing a gravitational field to exactly counter the centrifugal acceleration\cite{OzSchuCQG}. In any case, the inertial acceleration at a given point has only the Coriolis term.} There is a Coriolis-like inertial acceleration $2\omega v$ on a particle moving with non-relativistic speed $v$, perpendicular to the direction of motion\cite{OzSchuCQG}. This provides the centripetal acceleration $2v^2/d$ required  to keep the particle moving with speed $v$ on the circle of diameter $d$. Setting $2\omega v = 2v^2/d$ and we get $d=v/\omega$. Of course, as seen from the local inertial frames, the geodesic has zero acceleration and is locally `straight'.

Now consider the motion of a particle with constant speed $v\ll1$ along a circle of diameter $d\neq v/\omega$. Clearly the required centripetal acceleration $2v^2/d$ no longer equals the Coriolis acceleration $2\omega v$. An additional acceleration $A=2\omega v-2v^2/d$ in the radially outward direction would be required to maintain this trajectory at this speed. As seen from the local inertial frames, these worldlines would be accelerated and would have just this acceleration $A=2\omega v(1-v/\omega d)$. These worldlines have all the symmetry of the geodesics they have been derived from. All points on them are equivalent. They have a constant speed and an acceleration of constant magnitude, perpendicular to the velocity. All we need to do is extend this construction to the relativistic case.

This can be done by considerng the worldine of a particle W moving at constant speed $v$ along the trajectory
\begin{equation}\label{eqn:cylWorldlineTrajectory}
R = D \sin(\phi-\phi_0)
\end{equation}
in the $R$-$\phi$ plane, with $D$ specified independently of $v$. The worldline is a geodesic only if $v=D\sqrt{2}$. In general, each worldline of this 2-parameter family (labelled by $v$, $D$) has an acceleration, $A(v,D)$, which can be calculated neatly by comparing its worldline with that of a geodesic G of speed $v_g=D\sqrt{2}$ which projects to the same trajectory in the $R$-$\phi$ plane.

The 4-acceleration of the W worldline, $x^\rho(s)$, is
\begin{equation} \label{eqn:cylAccelerationDiffEqn}
A^\rho = \frac{d^2x^\rho}{ds^2} + \Gamma^{\rho}_{\mu\nu}\frac{dx^\mu}{ds} \frac{dx^\nu}{ds}
\end{equation}
while that of the geodesic G, $x_g^\rho(s)$, is
\begin{equation} \label{eqn:cylGeodesicEqn}
0 = \frac{d^2x_g^\rho}{ds_g^2} + \Gamma^{\rho}_{\mu\nu}\frac{dx_g^\mu}{ds_g} \frac{dx_g^\nu}{ds_g}
\end{equation}

The worldlines of W and G differ only in the $t$ and $s$ at which they reach a given point on the $R$-$\phi$ trajectory. So if we move along the two worldlines maintaining correspondence between points of same $(R,\phi)$, then $dR_g=dR$, $d\phi_g=d\phi$, the metric and $\Gamma^{\rho}_{\mu\nu}$ (which depend only on $R$) are the same at the corresponding points of W and G, but $ds_g$ and $dt_g$ differ from $ds$ and $dt$. (Subscript $g$ labels quantities for the geodesic G).

The latter can be related by rearranging the defintions of $v$ and $v_g$:
\begin{equation} \label{eqn:vdef}
v^2 = \frac{2(dR^2+R^2 d\phi^2)}{\left( (1-R^2)\omega dt+ 2R^2 d\phi \right)^2} 
= \frac{2(dR^2+R^2 d\phi^2)}{\left((1-R^2)\omega ds\right)^2 + 2(dR^2+R^2 d\phi^2)}
\end{equation}
\begin{equation} \label{eqn:vzerodef}
v_g^2 = \frac{2(dR^2+R^2 d\phi^2)}{\left( (1-R^2)\omega dt_g+ 2R^2 d\phi \right)^2} 
= \frac{2(dR^2+R^2 d\phi^2)}{\left((1-R^2)\omega ds_g\right)^2 + 2(dR^2+R^2 d\phi^2)}
\end{equation}
to the form
\begin{equation} \label{eqn:dsratio}
\frac{ds}{ds_g} = \frac{v_g}{v}\sqrt{\frac{1-v^2}{1-v_g^2}}
\end{equation}
\begin{equation} \label{eqn:dtdifference}
\frac{dt}{ds_g} - \frac{dt_g}{ds_g} = \left(\frac{v_g}{v}-1\right)\frac{1}{\sqrt{1-v_g^2}}
\end{equation}
Since $(ds/ds_g)$ is constant, $d^2x^\rho/ds^2=(ds_g/ds)^2 (d^2x^\rho/ds_g^2)$ and since $(dt/ds_g)-(dt_g/ds_g)$ is likewise a constant, while $R$ and $\phi$ are identical at corresponding points, $d^2x^\rho/ds_g^2=d^2x_g^\rho/ds_g^2$. Therefore using (\ref{eqn:cylGeodesicEqn}), (\ref{eqn:cylAccelerationDiffEqn}) reduces to
\begin{eqnarray} \label{eqn:CoriolisDiff}
A^\rho &=& \left( \frac{ds_g}{ds} \right)^2 
\left( \frac{d^2x_g^\rho}{ds_g^2} + \Gamma^{\rho}_{\mu\nu}\frac{dx^\mu}{ds_g} \frac{dx^\nu}{ds_g} \right) \nonumber \\
 &=& \left( \frac{ds_g}{ds} \right)^2 \Gamma^{\rho}_{\mu\nu}
\left( \frac{dx^\mu}{ds_g} \frac{dx^\nu}{ds_g} - \frac{dx_g^\mu}{ds_g} \frac{dx_g^\nu}{ds_g}\right)
\end{eqnarray}

Next, since $dx^\mu/ds_g$ differs from $dx_g^\mu/ds_g$ only in the $t$-component and $\Gamma^\rho_{00}=0$ for a stationary metric with constant $g_{00}$, only $\Gamma^\rho_{0i}\left( \frac{dt}{ds_g} \frac{dx^i}{ds_g} - \frac{dt_g}{ds_g} \frac{dx_g^i}{ds_g}\right)$ contributes, and in view of (\ref{eqn:dtdifference}) gives
\begin{equation}
A^\rho = \left( \frac{ds_g}{ds} \right)^2 \left( 2\Gamma^{\rho}_{0i}\frac{dx_g^i}{ds_g} \right) (\frac{v_g}{v}-1)\frac{1}{\sqrt{1-v_g^2}}
\end{equation}
Substituting into this, the only non-zero $\Gamma^{\rho}_{0i}$ viz. $\Gamma^0_{01}=2R/(1-R^2)$, $\Gamma^1_{02}=\omega R$ and $\Gamma^2_{01}= -\omega/R$, we get \footnote{The components are listed in the order $(t,R,\phi)$}
\begin{equation}
A^\rho = \frac{2\omega}{v}\left( \frac{ds_g}{ds} \right)^2 \frac{v_g-v}{\sqrt{1-v_g^2}}
\left( \frac{2R}{\omega(1-R^2)}\frac{dR}{ds_g},\ R\frac{d\phi}{ds_g},\ \frac{-1}{R}\frac{dR}{ds_g}\right) 
\end{equation}
Using (\ref{eqn:dsratio}) to substitute for $(ds_g/ds)$ and (\ref{eqn:vzerodef}) to replace the $ds_g$ in $(dR/ds_g)$ and $(d\phi/ds_g)$ in terms of $\sqrt{dR^2+R^2d\phi^2}$, simplifying the obtained fractions in $dR$ and $d\phi$ using $dR=d\phi\sqrt{D^2-R^2}$ and substituting $v_g=D\sqrt{2}$, we find that the contravariant components of the 4-acceleration are
\begin{equation}
A^\rho = A \left[ \frac{\omega(1-R^2)}{\sqrt{2}} \right]
\left( \frac{2R\sqrt{D^2-R^2}}{D\omega(1-R^2)},\ \frac{R}{D},\ -\frac{\sqrt{D^2-R^2}}{RD} \right)
\end{equation}
its covariant components are
\begin{equation}
A_\rho = A \left[ \frac{\sqrt{2}}{\omega(1-R^2)} \right]
\left( 0,\ -\frac{R}{D},\  \frac{R}{D}\sqrt{D^2-R^2} \right)
\end{equation}
and its magnitude is
\begin{equation} \label{eqn:Amag}
A = \sqrt{-A^\rho A_\rho} = \frac{2v\omega}{1-v^2} \left( 1-\frac{v}{D\sqrt{2}} \right)
\end{equation}
Thus the worldlines have a constant magnitude of acceleration $A$ determined entirely by the constant speed $v$ and the diameter, $D$, of the trajectory. In general $u^\mu A_\mu=0$, and in this case $A_0=0$. Thus $u^i A_i=0$, which implies that the 3-acceleration is perpendicular to the 3-velocity, as desired.

Note that A is signed. When $A>0$, the acceleration is opposite to the inertial Coriolis acceleration.\footnote{This means that the acceleration is `outwards' to the trajectory, but only as long as the acceleration is smaller than the Coriolis term. When the acceleration $A>0$ exceeds the Coriolis term, the sense of the trajectory reverses and the acceleration becomes inwards. These trajectories can be described by $D<0$ and are discussed in Section \ref{sec:cylWorldlineProperties}.} This can be verified most easily by taking the $v\ll1$ limit. We get $A\approx2\omega v(1-v/\omega d)$, where $d$ is the diameter of the trajectory in the $r$-$\phi$ plane (i.e. $\omega d/\sqrt{2}:=\tanh^{-1}D\approx D$). This result agrees perfectly with the non-relativistic arguments given earlier. 

The magnitude of the 4-acceleration, $A$, is equal to the magnitude of the 3-acceleration in a Local Inertial Frame (LIF) instantaneously at rest with W. Indeed, the thrust that a rocket of rest mass $m_0$ must generate to travel along the worldline W (as seen in its own rest frame) is $m_0 A$, directed perpendicular to its velocity with respect to the comoving frame. In particular, if this thrust is generated by ejecting gas/accelerated ions at relative speed $v_{rel}$, the rate of ejection of mass (in the rocket's rest frame) is given by
\begin{equation}\label{eqn:MassEjectionRate}
\left(-\frac{dm_0}{ds}\right) = \frac{\sqrt{1-v_{rel}^2}}{v_{rel}} m_0 A
\end{equation}

The LIF is however not meaningful for a photon and indeed $A$ diverges as $v\to1$.  \footnote{This is only to be expected because $A$ is the magnitude of $d^2 x^{\mu}/ds^2$, and $ds=0$ for $v=1$.} In this case, the required acceleration may be generated through forces applied by appropriately stationed objects comoving with the universe. The 3-force to be applied at any given point can be determined by transforming to the comoving LIF the 3-force, $F^{(3)}_{rest}=m_0 A$, we have been dealing with in the rest LIF of W .  This can, in turn, be done by applying to $m_0 A$ a Lorentz boost of speed $v$ perpendicular to $F^{(3)}_{rest}$. The result is 
\begin{eqnarray}
F^{(3)}_{comoving} &=& m_0 A \sqrt{1-v^2} = E A (1-v^2)  \nonumber\\
&=& 2v\omega E \left( 1-\frac{v}{D\sqrt{2}} \right)
\end{eqnarray}
where $E=m_0/\sqrt{1-v^2}$ is the total energy of the particle that is to be forced along a WSA of speed $v$ and diameter $D$. Note that $F^{(3)}_{comoving}$ is perfectly well-behaved and hence meaningful at $v=1$. As a concrete example, we could have an electromagnetic pulse ($v=1$) of energy $E$ relayed across a series ($N\gg1$) of comoving stations positioned along a circle of diameter $D$ on the $R$-$\phi$ plane.\footnote{With $D>D_{CTC}|_{v=1}=2\sqrt{2}/3$, this could be used for sending signals to the past.} $F^{(3)}_{comoving}$ is then approximately the impulse imparted to the photons at each station, divided by the transit time between successive stations. 

Before analysing these results further, we derive the expressions for the coordinate time $t$, and the proper time $s$ taken to get to a point on the worldline's trajectory in the $R$-$\phi$ plane.

\subsection{The $t$ and $s$ integrals} \label{sec:cylTIntegral}
We start by rearranging (\ref{eqn:vdef}), after eliminating $R$ from it using $R = D\sin(\phi-\phi_0)$, to the form
\begin{eqnarray}
\omega\frac{dt}{d\phi} = \frac{D\sqrt{2}/v - 2D^2\sin^2(\phi-\phi_0)}{1-D^2\sin^2(\phi-\phi_0)} \nonumber\\
\omega\frac{ds}{d\phi} = \frac{\sqrt{1-v^2}}{v} \frac{D\sqrt{2}}{1-D^2\sin^2(\phi-\phi_0)} 
\end{eqnarray}
These can be recast to
\begin{eqnarray} 
\omega(t-t_0) &=& \int^\phi_{\phi_0}  \left[ 2d\phi - (2 - D\sqrt{2}/v ) 
\frac{d\tan(\phi-\phi_0)}{1 + (1-D^2)\tan^2(\phi-\phi_0)} \right] \nonumber\\
\omega(s-s_0) &=& \frac{D\sqrt{2}}{v}\sqrt{1-v^2} \int^\phi_{\phi_0} \frac{d\tan(\phi-\phi_0)}{1 + (1-D^2)\tan^2(\phi-\phi_0)}
\end{eqnarray}
and readily integrated for $|D|<1$ to \footnote{Here the $\arctan$ has been defined with range $[0,\pi)$, and $\arctan(k \tan(\pi/2)) := \pi/2$. This makes this solution valid upto $\phi=\phi_0+\pi$, ie. up until the worldline has completed one orbit. The solution can be extended to any number of orbits, by adding $n\pi$ to the $\arctan$, where $n$ is the number of completed orbits.}
\begin{eqnarray} \label{eqn:cylTIntegralClosed}
\omega(t-t_0) &=& 2(\phi-\phi_0) - \frac{2-D\sqrt{2}/v}{\sqrt{1-D^2}} \arctan\left( \sqrt{1-D^2} \tan(\phi-\phi_0) \right) \nonumber\\
\omega(s-s_0) &=& \frac{\sqrt{1-v^2}}{v}\frac{D\sqrt{2}}{\sqrt{1-D^2}} \arctan\left( \sqrt{1-D^2} \tan(\phi-\phi_0) \right)
\end{eqnarray}
for $|D|=1$ to
\begin{eqnarray} \label{eqn:cylTIntegralCritical}
\omega(t-t_0) &=& 2(\phi-\phi_0) - (2-\sqrt{2}/v) \tan(\phi-\phi_0) \nonumber\\
\omega(s-s_0) &=& \frac{\sqrt{1-v^2}}{v} \sqrt{2}\tan(\phi-\phi_0)
\end{eqnarray}
and for $|D|>1$ to
\begin{eqnarray} \label{eqn:cylTIntegralOpen}
\omega(t-t_0) &=& 2(\phi-\phi_0) - \frac{2-D\sqrt{2}/v}{\sqrt{D^2-1}} \tanh^{-1}\left( \sqrt{D^2-1} \tan(\phi-\phi_0) \right) \nonumber\\
\omega(s-s_0) &=& \frac{\sqrt{1-v^2}}{v}\frac{D\sqrt{2}}{\sqrt{D^2-1}} \tanh^{-1}\left( \sqrt{D^2-1} \tan(\phi-\phi_0) \right)
\end{eqnarray}
Now that we have $t$, $R$ and $s$ as functions of $\phi$, we have a complete parametric representation of the worldlines. All the above results have been derived with a worldline with $D>0$ in mind. But the results are valid for $D<0$ as well, if we consider negative $D$ to correspond to trajectories of diameter $|D|$, but traversed in the opposite sense. \footnote{For the $D<0$ case however, $\phi$ decreases with increasing proper time, and the worldline that leaves the origin with $\phi=\phi_0$ returns to it with $\phi=\phi_0-\pi$. Hence the $\arctan$ used in (\ref{eqn:cylTIntegralClosed}) for $D<0$ must be defined with range $(-\pi,0]$, along with $\arctan(k \tan(-\pi/2)) := -\pi/2$}

\subsection{The Properties of the WSAs} \label{sec:cylWorldlineProperties}

\begin{figure}
\includegraphics[width=120mm]{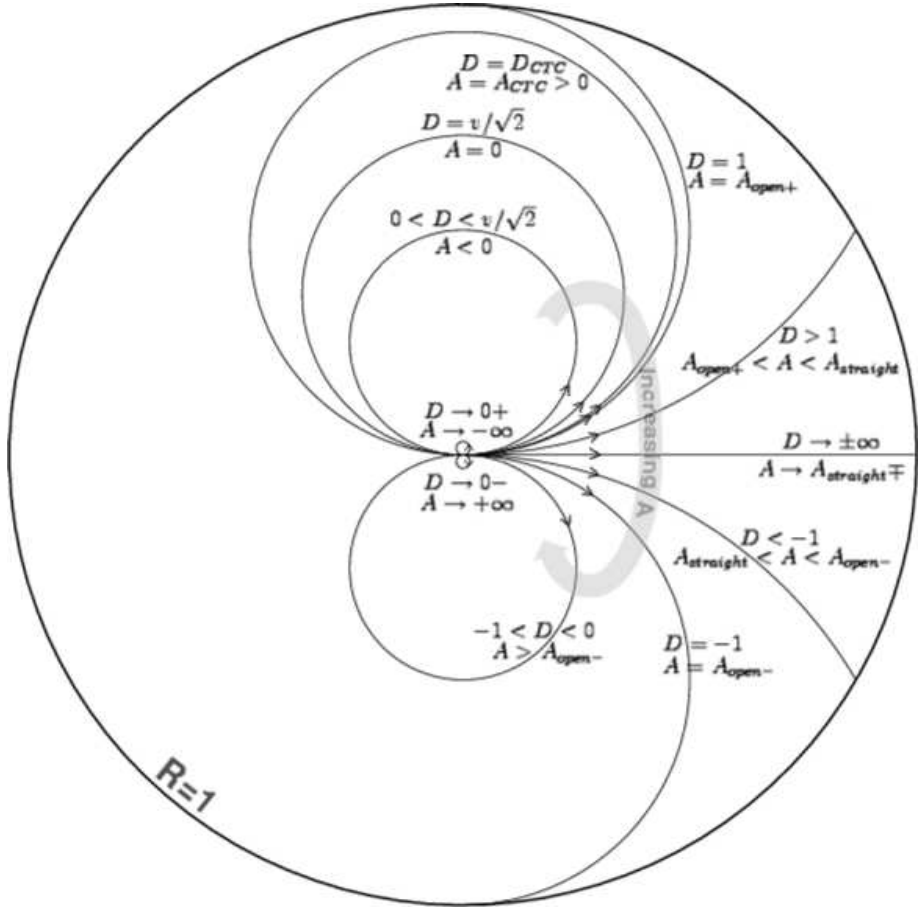}
\caption{WSA's of a given speed $v$. This diagram summarizes the relationship between $D$, $A$ and the projection of the trajectories of the correponding worldlines in the $R$-$\phi$ plane, for a given speed $v>1/\sqrt{2}$. (If $v<1/\sqrt{2}$, the diagram would be the same except that there would be no CTC, as $D_{CTC}>1$ for $v<1/\sqrt{2}$.)  }
\label{fig:WSAdiagram}
\end{figure}

As we have seen, WSAs depend on two parameters, $v\in[0,1]$ and $D\in(-\infty,\infty)$. $D$ is the diameter of the WSA's $R$-$\phi$-projection. It is positive for orbits corotating with geodesics, and negative for those counter-rotating. The broad behaviour of WSAs for a fixed $v$ and variable $D$ is depicted in Figure \ref{fig:WSAdiagram}.

Note that the geodesic of speed $v$ has a fixed diameter $D=v/\sqrt{2}$ in the $R$-$\phi$ plane and, of course, $A=0$. For an inward acceleration $A<0$, the applied force assists the inertial Coriolis acceleration and the trajectory shrinks to a $D<v/\sqrt{2}$. As the inward acceleration $A\to-\infty$, $D\to0$ and the trajectory shrinks to a point. For an outward acceleration $A>0$, the applied force opposes the Coriolis term, and makes $D>v/\sqrt{2}$. The trajectory continues to expand with increasing $A$, upto $A\to A_{open+}$, the critical `Escape Acceleration' 
\begin{equation}
A_{open+}=\frac{2v\omega}{1-v^2}\left(1-\frac{v}{\sqrt{2}}\right)
\end{equation}
At $A=A_{open+}$, $D=1$, and we get the critical escape trajectory, namely, a semicircle with the point at $R=1$ removed. As we increase the acceleration further, $D$ increases, and the trajectories turn into minor arcs starting from $R=0$ and terminating at $R=1$. The proper time taken to approach the $R=1$ point is infinite for all of them, as we would expect.

When the acceleration reaches the value
\begin{equation}
A_{straight}=\frac{2v\omega}{1-v^2}
\end{equation}
it exactly cancels the Coriolis term, and the trajectory becomes a straight radial line from $R=0$ to $R=1$. This corresponds to $D\to\infty$ (as also to $D\to-\infty$). Higher accelerations produce circular arcs curving in the opposite direction, and hence corresponding to $D<0$. Note that the acceleration is now inwards, while the Coriolis force is outwards. As $A$ increases beyond $A_{straight}$, $D$ starts increasing from $-\infty$, and the trajectories are circular arcs of the opposite sense terminating at $R=1$. This continues till $A$ reaches the value
\begin{equation}
A_{open-} = \frac{2v\omega}{1-v^2} \left(1+\frac{v}{\sqrt{2}}\right)
\end{equation}
for which $D=-1$. This corresponds to the critical escape trajectory in the opposite sense --- likewise a semicircle terminating at $R=1$. Thus for $A \in [A_{open+}, A_{open-}]$, we get $|D|\ge1$ and hence, open (unbounded) trajectories.

For accelerations higher than $A_{open-}$, the trajectories are circles in the $R$-$\phi$ plane of diameters $D\in(-1,0)$, directed oppositely to the spacetime's geodesics.

The proper and coordinate times required to circumnavigate the closed orbits identified above have a series of interesting properties. For $D<1$, the worldline which leaves the origin along $\phi=\phi_0$, returns to it along $\phi=\phi_0+\pi$. Setting $\phi=\phi_0+\pi$ in (\ref{eqn:cylTIntegralClosed}) we obtain times $\Delta t$ and $\Delta s$ required to complete the orbit, as measured by the observer O stationed at the origin, and the observer W traveling along the worldline, respectively.
\begin{equation} \label{eqn:cylDeltaTorbit}
\Delta t_{round-trip}^{D>0} = \frac{\pi}{\omega} \left( 2 - \frac{2-D\sqrt{2}/v}{\sqrt{1-D^2}} \right)
\end{equation}
\begin{equation} \label{eqn:cylDeltaSorbit}
\Delta s_{round-trip}^{D>0} = \frac{\pi}{\omega} \left( \sqrt{1-v^2}\frac{D\sqrt{2}/v}{\sqrt{1-D^2}} \right)
\end{equation}
For geodesics, $D = v/\sqrt{2}$ and the above expressions reduce to
\begin{eqnarray} \label{eqn:cylGeodesicOrbitTime}
\Delta t_{round-trip}^{geodesic} &=& \frac{\pi}{\omega} \left( 2 - \frac{1}{\sqrt{1-v^2/2}}\right) \nonumber\\
\Delta s_{round-trip}^{geodesic} &=& \frac{\pi}{\omega} \frac{\sqrt{1-v^2}}{\sqrt{1-v^2/2}}.
\end{eqnarray}
For $v\to0$, both O and W observe the time for the round trip to be $\pi/\omega$. Geodesic round-trips of higher $v$ complete sooner, according to both the observers. For null geodesics, that is at $v=1$, the time observed by W is $0$ as expected, while that observed by O is $(\pi/\omega) (2-\sqrt{2})$. 

For a fixed $v$, $\Delta s_{round-trip}$ is a monotonically increasing function of $D$, which means that spatially larger orbits take longer to complete, as observed by the traveler W. At small $D$, $\Delta s_{round-trip}$ varies linearly with $D$ and as $D\to1$, $\Delta s_{round-trip}\to\infty$ which simply means that orbits of infinite spatial extent take infinite proper time to complete. Hence the behaviour of $\Delta s_{round-trip}$ is perfectly normal.

By contrast, the behaviour of $\Delta t_{round-trip}$ is completely normal only at small $D$ where it varies linearly with $D$. Indeed, $\Delta t_{round-trip}$ is not monotonic in $D$ for fixed $v$. It has a maximum at $D=1/v\sqrt{2}$. As long as $v<1/\sqrt{2}$, it does not behave strangely for any $D\in(0,1)$ and as $D\to1$, $\Delta t_{round-trip}\to+\infty$. But for $v>1/\sqrt{2}$, as $D\to1$, $\Delta t_{round-trip}\to-\infty$! This means that by moving along a sufficiently large orbit at a speed greater than $1/\sqrt{2}$, W can travel as far into the past (as measured by O) as desired. In particular, W can choose to move along a Closed Time-like Curve for which $\Delta t_{round-trip}=0$ and hence 
\begin{equation} \label{eqn:DAforCTC}
D_{CTC}=\frac{2\sqrt{2}v}{1+2v^2}\ \textrm{ ie. }\ 
A_{CTC}=\frac{v\omega}{2}\left(\frac{3-2v^2}{1-v^2}\right)
\end{equation}
For any $v>1/\sqrt{2}$, worldlines with $D>D_{CTC}$ or equivalently $A>A_{CTC}$ are Past-Travelling Worldlines. Of course $D$ must be $<1$ for the orbit to be closed in the first place.\footnote{From the viewpoint of fuel efficiency, the optimum CTC corresponds to $v_{opt}=\sqrt{\sqrt{3}/2}\approx0.9306$. Indeed, by (\ref{eqn:MassEjectionRate}), the mass ejected over the round-trip is $\Delta m = m^i_0 \left(1-\exp\left(-A\Delta s_{round-trip}\sqrt{1-v^2_{rel}}/v_{rel}\right)\right)$. Clearly, the minimum of $\Delta m$ coincides with that $A \Delta s_{round-trip}$. $v_{opt}$ minimises the latter expressed as a function of $v$ alone using (\ref{eqn:Amag}), (\ref{eqn:cylDeltaSorbit}) and (\ref{eqn:DAforCTC}).}

As noted above, there is a second set of closed orbits, directed oppositely to the space-time's geodesics. For this set, the higher the acceleration, the smaller the diameter, to the point that $D\to0-$ as $A\to+\infty$. We can calculate the time taken to complete any of these orbits  exactly as in section \ref{sec:cylTIntegral}. The only changes are that $D<0$ and our integral over $\phi$ goes from $\phi_0$ to $\phi_0-\pi$, since $\phi$ now decreases with increasing proper time. The results are:
\begin{equation} \label{eqn:cylDeltaTorbitReverse}
\Delta t_{round-trip}^{D<0} = \frac{\pi}{\omega} \left( \frac{2-D\sqrt{2}/v}{\sqrt{1-D^2}} - 2 \right)
\end{equation}
\begin{equation} \label{eqn:cylDeltaSorbitReverse}
\Delta s_{round-trip}^{D<0} = -\frac{\pi}{\omega} \left( \sqrt{1-v^2}\frac{D\sqrt{2}/v}{\sqrt{1-D^2}} \right)
\end{equation}
In fact, these $D<0$ formulae can be obtained from their $D>0$ counterparts, (\ref{eqn:cylDeltaTorbit}) and (\ref{eqn:cylDeltaSorbit}), simply by replacing $s\to-s$ and $t\to-t$. Further, for these orbits
$$\left(\Delta t_{round-trip}^{D<0}\right) = \frac{1}{\sqrt{1-v^2}}\left(\Delta s_{round-trip}^{D<0}\right)
 + \frac{2\pi}{\omega}\left( \frac{1}{\sqrt{1-D^2}}-1 \right) $$
Hence $\left(\Delta t_{round-trip}^{D<0}\right) \sqrt{1-v^2} > \left(\Delta s_{round-trip}^{D<0}\right) > 0$. In particular, there are no CTCs of the reverse sense. Moreover, roughly speaking, the curvature effects aid the time-dilation effect i.e. W can `travel further into the future' of O than possible in a flat spacetime, by making a round trip at a given speed $v$.

\begin{figure}
\includegraphics[width=120mm]{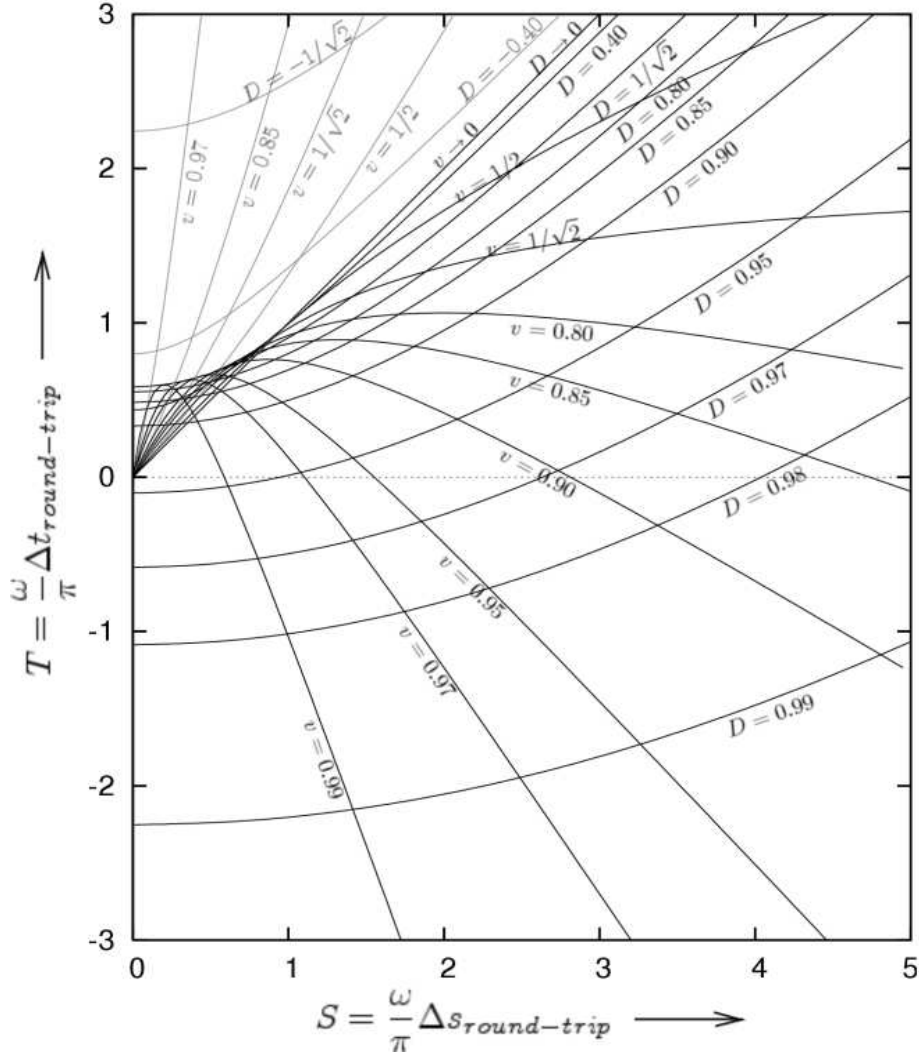}
\caption{Lines of constant $v$ and constant $D$ on a $T$ vs $S$ plot. Each set of lines can be visualised as a grid drawn on a stretched sheet. There are two distinct sheets, one for $D\in(-1,0)$ which stretches over the entire region $T>S>0$ (drawn in light gray). The other sheet for $D\in(0,1)$ (drawn in black) has a clear fold along the arc $T=2-\sqrt{2-S^2}$, $S\in(0,1)$. A single fold of this sheet covers the entire $T<S$, $S>0$ region, while two folds of this sheet cover the region between the line $T=S$ and the arc $T=2-\sqrt{2-S^2}$, $S\in(0,1)$. First look at the $D>0$ sheet. The lines of constant $v$ for $v<1/\sqrt{2}$ always have positive slope, and $T$ always increases with increasing $S$. However for $v>1/\sqrt{2}$, there is a maximum beyond which $T$ begins to decrease with increasing $S$. There is a finite $S$ at which $T=0$, and for $S\to +\infty$, $T\to -\infty$. The lines of constant $D$ always have positive slope - at fixed $D$ if traveler W observes a longer time, so will the base station O. For $D>1/\sqrt{2}$, the lines of higher $D$ lie entirely below lines of smaller $D$. So once W crosses the G\"odel Horizon, the further W goes keeping $S$ fixed, the smaller is $T$. In the $D<0$ sheet, both $T$ and $S$ increase monotonically with increasing $v$ as well as with increasing $|D|$.}
\label{fig:TSPlot}
\end{figure}

All these results are best summarized by a $\Delta t_{round-trip}$-$\Delta s_{round-trip}$ diagram (See Figure \ref{fig:TSPlot}), which we draw for convenience in terms of the dimensionless round-trip times - $T:=\omega\Delta t_{round-trip}/\pi$ and $S:=\omega\Delta s_{round-trip}/\pi$. 

There are 3 distinct regions visible on this $T$-$S$ plane, differing in the number and nature of solutions of $(v,D)$ for each point $(T,S)$ in them. \footnote{The solutions $(v,D)$ for each $(T,S)$ can be obtained by inverting (\ref{eqn:cylDeltaTorbit}), (\ref{eqn:cylDeltaSorbit}) and (\ref{eqn:cylDeltaTorbitReverse}), (\ref{eqn:cylDeltaSorbitReverse}). The equations are quadratics, and the solutions must be checked by back-substituting, to discard spurious solutions.}

The first region, $S\in(0,\infty)$ and $T\in(-\infty,S)$, is bounded by the line $T=S$, which corresponds to the non-relativistic limit, $v\to0$ and $D\to0$. In this region, there is exactly one value of $v$ and $D$, with $D>0$, for each value of $T$ and $S$ namely:
$$D=\sqrt{1-\frac{1}{F^2}}\textrm{, }\ v=\left(1+\frac{S^2}{2(F^2-1)}\right)^{-1/2}$$ where
$$F = F(S,T) = (2-T) + \sqrt{\frac{1}{2} \left[ (2-T)^2+S^2-2 \right] } $$
Note that W can travel as far back into the past of O as desired, with as small a proper time lapse as desired, by making an appropriate choice of $v$ and $D$ (that is, bringing them both sufficiently close to $1$).

The second region corresponds to $S\in(0,1)$ and $T\in(S,2-\sqrt{2-S^2})$, lying between the above mentioned line and a circle of radius $\sqrt{2}$ centred at $T=2$, $S=0$. In this region, each value of $T$ and $S$ gives rise to exactly three solutions for $v$ and $D$. Two of the solutions correspond to $D>0$, and are given exactly as in the first region, except that in this case there are two possibilities for $F$:
$$F = F_{\pm}(S,T) = (2-T) \pm \sqrt{\frac{1}{2} \left[ (2-T)^2+S^2-2 \right] } $$
The third solution corresponds to $D<0$, and is given by:
$$D=-\sqrt{1-\frac{1}{G^2}}\textrm{, }\ v=\left(1+\frac{S^2}{2(G^2-1)}\right)^{-1/2}$$ where
$$G = G(S,T) = (2+T) - \sqrt{\frac{1}{2} \left[ (2+T)^2+S^2-2 \right] } $$

The rest of the $T$-$S$ plane constitutes the third region, where there is exactly one solution for $v$ and $D$, with $D<0$. This solution is exactly the same as the $D<0$ solution in the second region.

These results can be easily visualised by imagining the iso-$v$ and iso-$D$ lines on the diagram as two sheets, a sheet for $D<0$ and a `folded' sheet for $D>0$. There are as many solutions of $(v,D)$ at any $(T,S)$ point as the total number of overlapping surfaces of both these sheets at that point.

We end this section by pointing out an interesting property of low velocity geodesics. First note that
$$\lim_{D\to0} \frac{\Delta s_{round-trip}}{\Delta t_{round-trip}} = \sqrt{1-v^2} $$
This is just the usual time dilation effect in a twin-paradox-like scenario, as the effect of the spacetime's curvature vanishes in the limit $D\to0$. The `traveller' W measures a time which is smaller by a factor of $\sqrt{1-v^2}$ than the time measured by the `stationary' observer O. Note that in this case O is moving geodesically, while W is accelerated.

Contrast this with a scenario where W is also moving geodesically, but with a non-relativistic speed $v\ll 1$. Using (\ref{eqn:cylGeodesicOrbitTime}), we see that
$$\frac{\Delta s_{geodesic}}{\Delta t_{geodesic}} = \frac{\sqrt{1-v^2}}{2\sqrt{1-v^2/2}-1} = 1 + O(v^4) $$
Since W makes a round trip in this case and returns to the base station O, without either of them ever accelerating, they are both on an equal footing. Indeed they measure the same time for the journey upto $O(v^2)$. At this order, the effect of the curvature seems to be cancelling the effect of time dilation!

\subsection{Visualizations} \label{sec:cylVisuals}

\begin{figure}
\includegraphics[width=120mm]{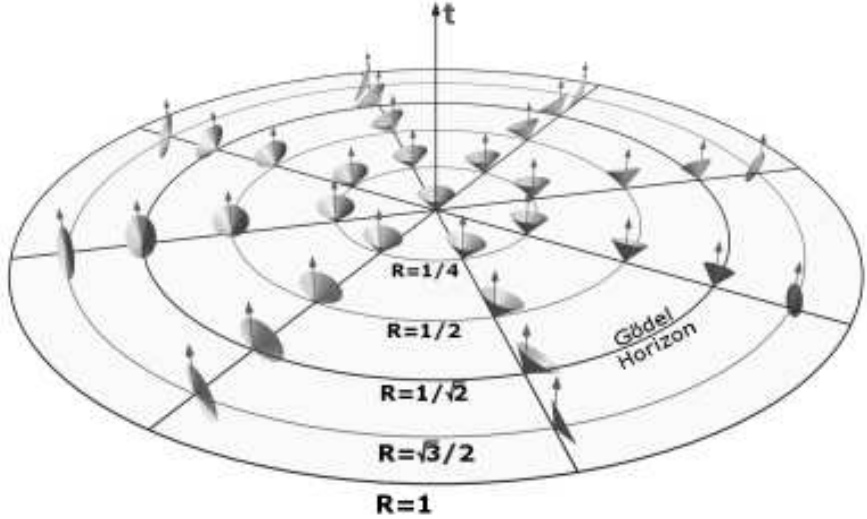}
\caption{A map of the future lightcones of the G\"odel universe in the cylindrical $(t, R, \phi)$ $0 \le R < 1$ coordinates. All the light-cones contain the future-directed $t$-axis. Note that the G\"{o}del horizon is a CNC.}
\label{fig:cylLightConeDiagram}
\end{figure}

We further understand these results through sketches containing representative light-cones of the spacetime.  The light-cone structure of the G\"odel universe as a whole in the {\it compact} $R$-$\phi$ plane (see Figure \ref{fig:cylLightConeDiagram}) is much the same as its light-cone structure of the infinite $r$-$\phi$ plane \cite{HawkingEllis}.

\begin{figure}
\includegraphics[width=120mm]{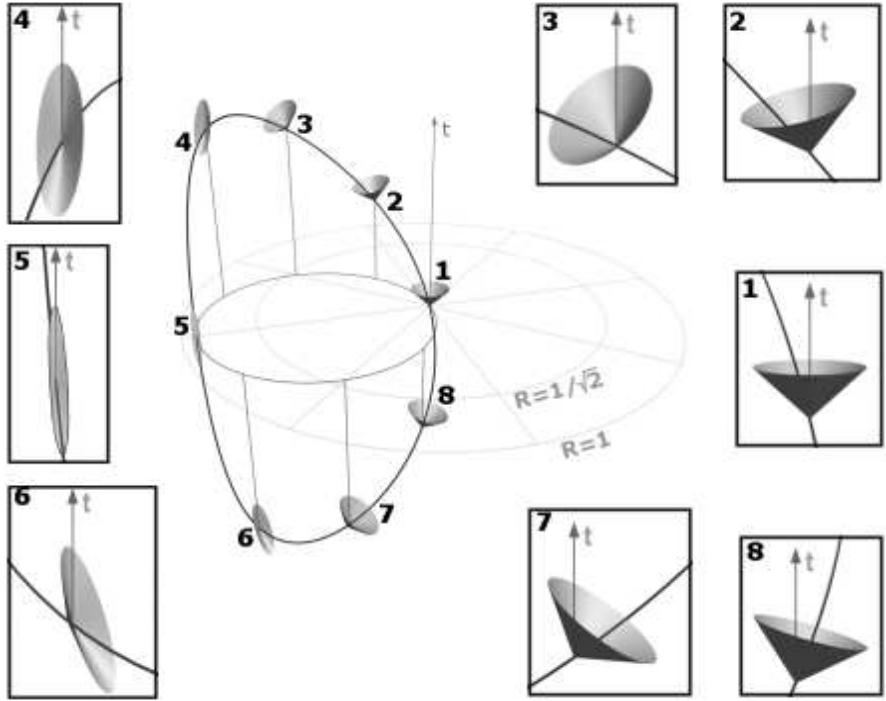}
\caption{A diagram depicting a Closed Null Curve passing through the origin from an interesting perspective. The tangent of the worldline rotates on the light-cone, and the light-cones beyond the G\"{o}del Horizon have a past-directed section as well.}
\label{fig:cylDetailedCNC}
\end{figure}

Here too, a CNC forms at the G\"odel horizon, but once again, does not give much insight into how CNCs form in general. This insight is provided by Figure \ref{fig:cylDetailedCNC}, which graphically depicts  how the opening and tilting of light-cones allow a generic CNC passing through the origin to close. The way this works is this: Light-cones beyond the G\"odel horizon have sections pointing towards the past and both $dt$ and $d\phi$ are time-like in this region. Since $dt$ is negative when $d\phi$ is positive and {\it vice versa}, the signature of the metric is maintained: $t$ moves back locally but time does not, because we simultaneously move forward in $\phi$ which is also time-like. In what ratio we move back in one and forth in the other depends on where in $R$ we are. By going to an appropriately large value of $D$ ($D>D_{CTC}$, to be precise) the variation in $R$ over the orbit becomes sufficiently large that, globally, we end up moving back in both $t$ and $\phi$. This is where acausal behaviour sets in. 

Additional insights into the emergence of CNCs/CTCs can be gleaned by generalizing the portrayal by Hawking and Ellis of the refocussing of photons to WSAs. 

\begin{figure}
\includegraphics[width=120mm]{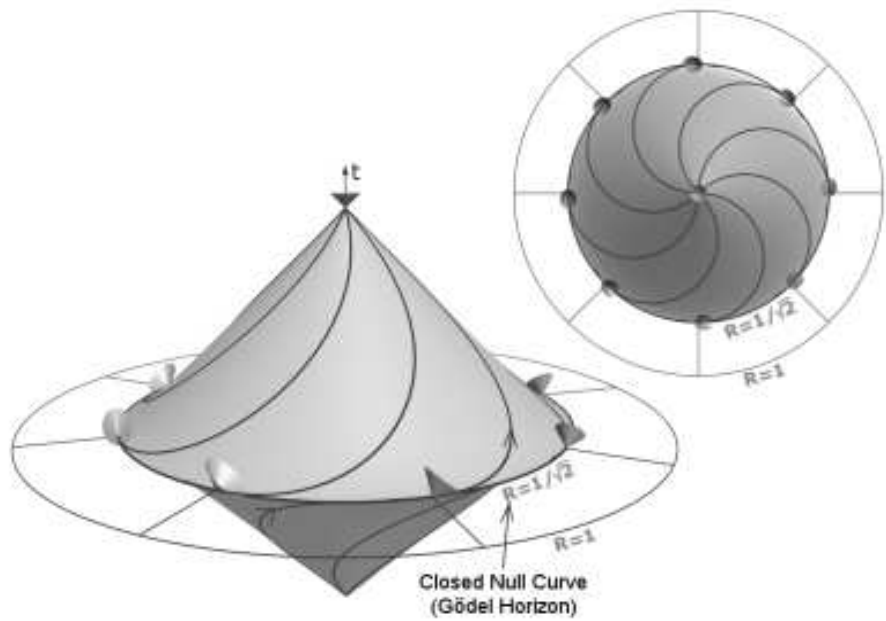}
\caption{The Hawking-Ellis picture in $(R,\phi)$ space: Envelopes of the null geodesics starting from the origin at the same $t$ in different initial directions $\phi_0$. A cusp develops at the G\"odel horizon, which itself forms a Closed Null Curve. This diagram should be compared to its counterpart, Figure \ref{fig:kundtNullGeodesicFamily}, in Kundt's coordinates.}
\label{fig:cylNullGeodesicFamily}
\end{figure}

We start by reproducing, in Figure \ref{fig:cylNullGeodesicFamily}, the Hawking and Ellis picture for $R\phi$ space. Since spatial infinity is brought to a finite value, this space is somewhat easier to both depict and visualise.

\begin{figure}
\includegraphics[width=120mm]{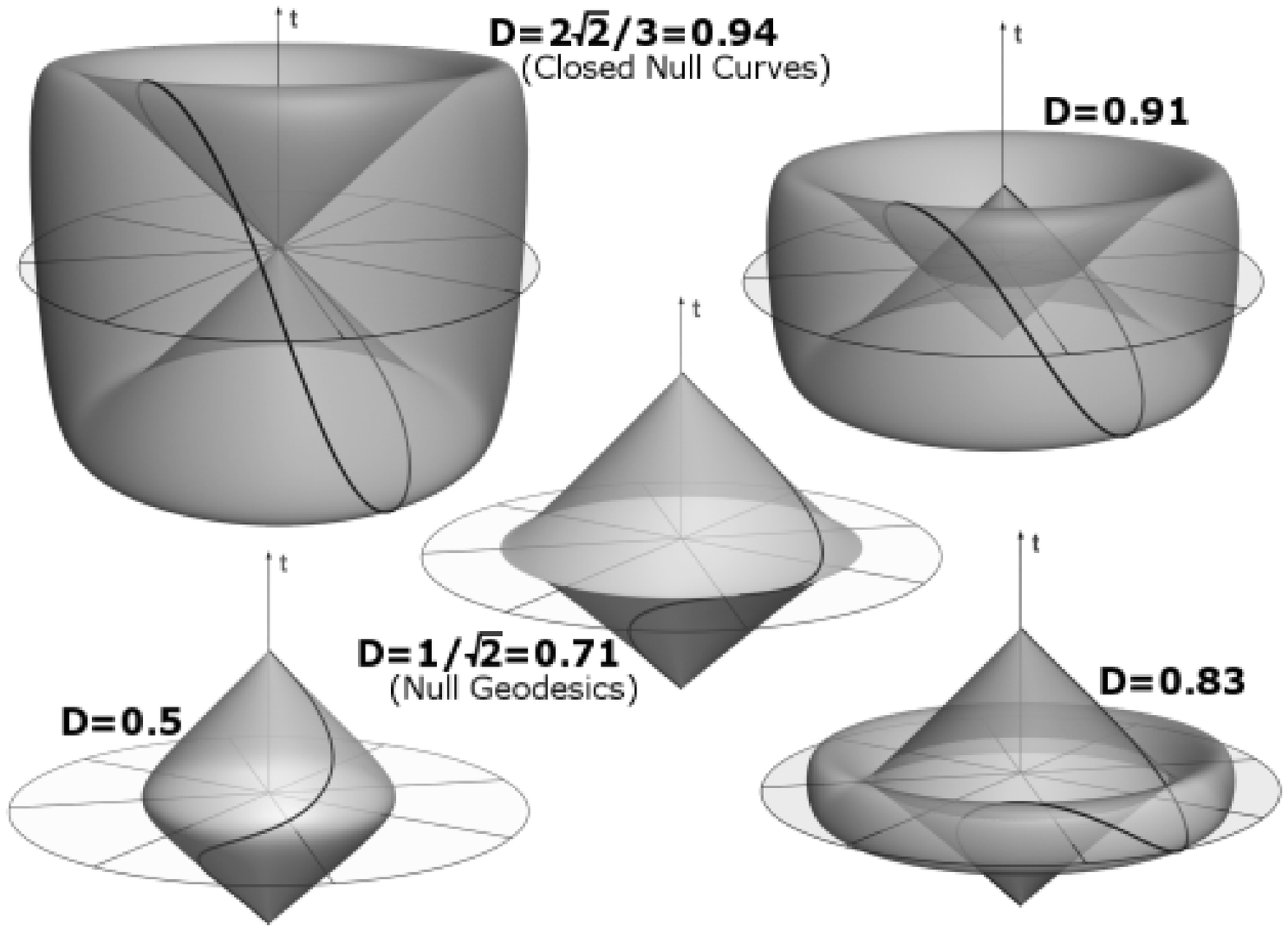}
\caption{Envelopes of all the $v=1$ worldlines emanating from the origin at the same $t$ in different initial directions $\phi_0$, for various fixed values of $D$ (or, equivalently, of $A$). The photons refocus at the origin. For $D=1/\sqrt{2}$, that is, for null geodesics, the envelope has a cusp at the G\"{o}del horizon. Note that the apex of the (lower) cone of divergence approaches that of the (upper) cone of reconvergence as $D$ increases. For $D=2\sqrt{2}/3$, the value corresponding to CNCs, the photons reconverge at the same point in {\it both} space and time. On each of the 5 envelopes, we display a representative (null) worldline.}
\label{fig:cylCNCEnvelope}
\end{figure}

Since WSAs have all the symmetries of geodesics, WSAs diverging from the origin with identical values of $v$ and $D$ (or equivalently, $A$) will refocus after a definite period, $T(v, D)$.  In Figure \ref{fig:cylCNCEnvelope}, we examine this refocussing for WSAs with $v=1$ and various values of $D$ (including $D=D_{Null Geodisic}$ and $D=D_{CTC}$) by drawing the envelopes of the diverging and reconverging worldlines.

\begin{figure}
\includegraphics[width=120mm]{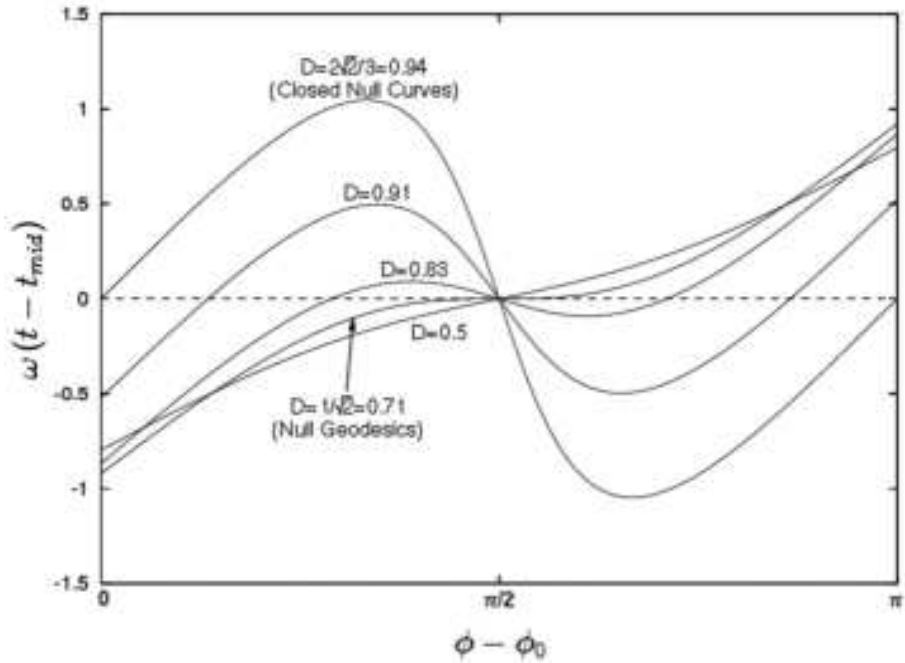}
\caption{ The $t(\phi)$-plots of $v=1$ worldlines for various values of $D$. (Note: $t_{mid}$ is the values of $t$ at $\phi=\phi_0+\frac{\pi}{2}$). To get the curve in $(t, R, \phi)$-space, curl up the $\phi$-axis, aligning the $\phi-\phi_0 = 0, 1$ lines with each other, place the cylinder so obtained with $\phi-\phi_0=0$ along the $t$-axis at the origin, and rescale the cylinder's diameter to $D$. Note that $\Delta t=t(\pi)-t(0)$ is a maximum for $D=1/\sqrt{2}$, {\it i.e.} for null geodesics. Thus acceleration, whatever be its sign, brings the photon back earlier to the base-station. This interesting property does not carry over to material particles (because $D_{geodesic}=v/\sqrt{2}$, whereas, $D_{max\Delta T}=1/v\sqrt{2}$, and the two are the same only for $v=1$).}
\label{fig:cylTimePhiCurves}
\end{figure}

We note for $D=D_{CTC}$, the WSAs refocus at  the space-time event of their departure. In each of the envelopes, we have traced a representative WSA.  In Figure \ref{fig:cylTimePhiCurves}, we cut open out and flatten out the cylinders on which these representatives lie. 

We note that $t(\phi)$ is monotonic only for the $D=0.5$ and $1/\sqrt{2}$ curves, which lie entirely within the G\"odel horizon. For larger values of $D$, the slope of $t(\phi)$ is in part negative. As explained above, this does not mean that we are moving backward in time here. Indeed, we are moving forward in $\phi$ which is now likewise time-like.

\section{Kinematics - Rectangular Coordinates} \label{sec:kundtKinematics} 

In the previous section, we solved the geodesic equation only for geodesics that pass through the origin because the solutions for geodesics that do not are in general much `dirtier', As a consequence, the accelerated worldlines to which we extended the geodesics passed through the origin as well. Since the translational symmetry of the G\"odel universe is not manifest in the cylindrical coordinate system, we could not generalize our results straightforwardly to arbitrary geodesics and WSAs. 

In this section, we accordingly migrate our results to the rectangular coordinate system $(x^0,x,y)$ with $y>0$, introduced by Kundt \cite{KundtOrigPaper}. This has a manifest translational symmetry, which we can use to free our worldlines of the need to pass through any fixed point.  The line element is:
\begin{eqnarray} \label{eqn:kundtLineElement}
ds^2 = \left( dx^0+\frac{dx}{\omega y} \right)^2 -  \left(\frac{dx^2+dy^2}{2\omega^2 y^2}\right)
\end{eqnarray}

The transformation relations between the rectangular and  cylindrical coordinates are
\begin{eqnarray} 
\omega x &=& \frac{2R \sin \phi}{1+R^2 + 2R \cos \phi} \label{eqn:coordtransformx}\\
\omega y &=& \frac{1-R^2}{1+R^2 + 2R \cos \phi} \label{eqn:coordtransformy}\\
\omega x^0 &=& \omega t + 2 \arctan \left(\frac{1-R}{1+R}\tan\frac{\phi}{2}\right) - \phi \label{eqn:coordtransformt}
\end{eqnarray}

\begin{figure}
\begin{center}\parbox{80mm}{\includegraphics[width=80mm]{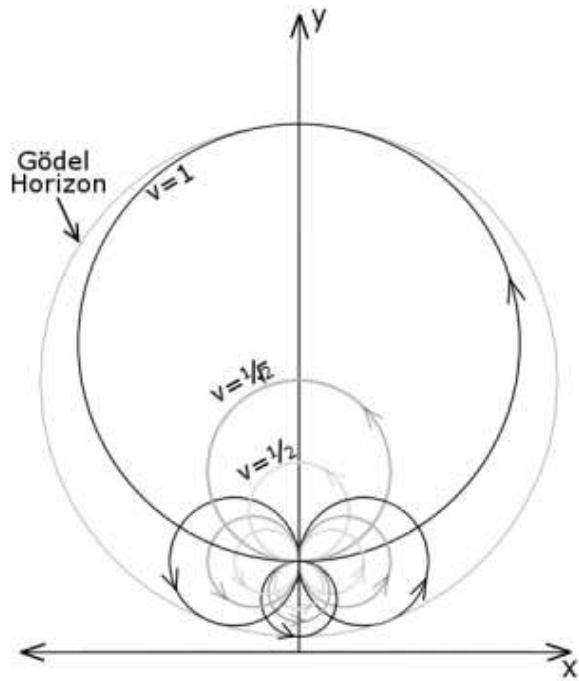}}\end{center}
\caption{The $x$-$y$ projections of the geodesics of $v=1,\frac{1}{\sqrt{2}}\ \textrm{and}\ \frac{1}{2}$ diverging from and refocussing at $(x,y)=(0,\omega^{-1})$, which corresponds to the origin of the cylindrical coordinates. The G\"odel Horizon corresponding to this point is shown. Note that particles at the same speed leaving $(0,\omega^{-1})$ simultaneously in different directions describe circles of different radii, but return at the same instant of time. This refocussing is a consequence of the underlying rotational symmetry.  }
\label{fig:kundtGeodesics}
\end{figure}

The $xy$-projections of the WSAs can be obtained in rectangular coordinates by eliminating $R$ and $\phi$ from $R=D\sin(\phi-\phi_0)$ using (\ref{eqn:coordtransformx}) and (\ref{eqn:coordtransformy}). The result is
\begin{equation} \label{eqn:kundtTrajectoryOrigin}
(x-x_c)^2+(y-y_c)^2=(y_c D)^2
\end{equation}
with
\begin{equation}
\omega x_c = \frac{D\cos\phi_0}{1-D\sin\phi_0},\ \ \omega y_c = \frac{1}{1-D\sin\phi_0}
\end{equation}
We see that the circular projections of diameter $D$ passing through the origin in the $R$-$\phi$ plane are likewise circles in the $xy$ plane centred at $(x_c, y_c)$, with radius $y_c D$. The values of $x_c$ and $y_c$ are determined completely by $D$ and $\phi_0$. All these trajectories pass through the point $x=0$, $y=\omega^{-1}$ which corresponds to the origin $R=0$ in cylindrical coordinates. However, since the radius, $y_c D$ is $\phi_0$-dependent, geodesics emanating in different directions from $(0, \omega^{-1})$ with the same $D$ describe circles of different radii (see Figure \ref{fig:kundtGeodesics}).

To free these trajectories from having to pass through the origin, we simply use the most general translation in the plane of rotation of the G\"odel universe $(x, y, x^0)\rightarrow (a x+b, a y, x^0)$ with $a>0$, which leaves the metric form-invariant, and hence the dynamics unchanged. Under this transformation, the worldline continues to be given by (\ref{eqn:kundtTrajectoryOrigin}), except that $x_c\rightarrow a x_c +b$ and $y_c\rightarrow a y_c$. By an appropriate choice of $a$ and $b$, $x_c$ can be made to take any value in $(-\infty,\infty)$ and $y_c$ any value in $(0,\infty)$.  Thus the most general worldline of constant acceleration and constant speed can be written parametrically as
\begin{eqnarray} \label{eqn:kundtTrajectory}
x &=& x_c + (y_c D) \cos\theta \nonumber\\
y &=& y_c + (y_c D) \sin\theta 
\end{eqnarray}
for any $x_c\in\mathcal{R}$ and $y_c>0$! 

Note further that the transformations between $x$,$y$ and $R$,$\phi$ do not involve $t$ or $x^0$. This implies that locally, the transformations between $(\partial_t, \partial_R, \partial_\phi)$ and $(\partial_{x^0}, \partial_x, \partial_y)$ are free of boosts. Hence the speed of a particle is identical as observed by comoving observers of both coordinate systems. In particular, the equation of the geodesic of speed $v$ is (\ref{eqn:kundtTrajectory}) with $D=v/\sqrt{2}$. The magnitude of acceleration is in any case a 4-scalar, and hence invariant. Thus the acceleration of a worldline of speed $v$ with trajectory (\ref{eqn:kundtTrajectory}) for a particular value of $D$ is given by exactly the same relation as in cylindrical coordinates:
\begin{equation}
A = \frac{2v\omega}{1-v^2} \left( 1-\frac{v}{D\sqrt{2}} \right)
\end{equation}
Using the differentials of relations (\ref{eqn:coordtransformx}), (\ref{eqn:coordtransformy}) and  (\ref{eqn:coordtransformt}), we can transform the components of the acceleration in cylindrical corrdinates to those in rectangular coordinates. The covariant components of the accaleration at the point $\theta$ on the worldline (\ref{eqn:kundtTrajectory}) are then
\begin{equation}
A_\rho = \frac{A}{\omega y \sqrt{2}} \left( 0,\ -\cos\theta,\  -\sin\theta \right)
\end{equation}
and the contravariant components are
\begin{equation}
A^\rho = A\omega y \sqrt{2} \left( \frac{-\cos\theta}{\omega y},\ \cos\theta,\  \sin\theta \right)
\end{equation}

\subsection{The $x^0$ and $s$ integrals} \label{sec:kundtTIntegral}
To complete the description of the worldlines in the rectangular coordinates, we need to relate $x^0$ and $s$ to the trajectory parameter $\theta$. The speed as given by (\ref{eqn:speeddefinition}) is:
\begin{equation}
v^2 = \frac{dx^2+dy^2}{2(\omega y dx^0 + dx)^2}
= \frac{dx^2+dy^2}{2(\omega y ds)^2 + (dx^2+dy^2)}
\end{equation}
Using (\ref{eqn:kundtTrajectory}), these can be rearranged into differential equations for $x^0(\theta)$ and $s(\theta)$:
\begin{eqnarray}
\omega\frac{dx^0}{d\theta} &=& \frac{D(\sin\theta + 1/v\sqrt{2})}{1+D\sin\theta} \\
\omega\frac{ds}{d\theta} &=& \frac{\sqrt{1-v^2}}{v\sqrt{2}} \frac{D}{1+D\sin\theta}
\end{eqnarray}
which can be recast as
\begin{eqnarray}
\omega(x^0-x^0_0) &=& \int^\theta_{\theta_0}  \left[ d\theta - (2 - D\sqrt{2}/v ) 
\frac{d\tan\frac{\theta}{2}}{(1-D^2)+(D+\tan\frac{\theta}{2})^2} \right] \nonumber\\
\omega(s-s_0) &=& \frac{D\sqrt{2}}{v}\sqrt{1-v^2} \int^\theta_{\theta_0} \frac{d\tan\frac{\theta}{2}}{(1-D^2)+(D+\tan\frac{\theta}{2})^2}
\end{eqnarray}
and readily integrated for $|D|<1$ to\footnote{In this case, the $\arctan$ has been defined with range $[\frac{\theta_0}{2},\ \frac{\theta_0}{2}+\pi)$. This solution is valid for one orbit ie. for $\theta\in[\theta_0,\ \theta_0+2\pi)$. To extend the solution to any number of orbits, add $n\pi$ to the $\arctan$, where $n$ is the number of completed orbits.}
\begin{eqnarray} \label{eqn:rectTIntegralClosed}
\omega(x^0-x^0_0) &=& (\theta-\theta_0) - \frac{2-D\sqrt{2}/v}{\sqrt{1-D^2}}
\left[ \arctan\left(\frac{D+\tan\frac{\theta}{2}}{\sqrt{1-D^2}}\right) \right]^\theta_{\theta_0} \nonumber\\
\omega(s-s_0) &=& \frac{\sqrt{1-v^2}}{v}\frac{D\sqrt{2}}{\sqrt{1-D^2}}
\left[ \arctan\left(\frac{D+\tan\frac{\theta}{2}}{\sqrt{1-D^2}}\right) \right]^\theta_{\theta_0}
\end{eqnarray}
for $|D|=1$ to
\begin{eqnarray} \label{eqn:rectTIntegralCritical}
\omega(x^0-x^0_0) &=& (\theta-\theta_0) - (2-\sqrt{2}/v)
\left[ \frac{-1}{D+\tan\frac{\theta}{2}} \right]^\theta_{\theta_0} \nonumber\\
\omega(s-s_0) &=& \frac{\sqrt{1-v^2}}{v} \sqrt{2}
\left[ \frac{-1}{D+\tan\frac{\theta}{2}} \right]^\theta_{\theta_0}
\end{eqnarray}
and for $|D|>1$ to
\begin{eqnarray} \label{eqn:rectTIntegralOpen}
\omega(x^0-x^0_0) &=& (\theta-\theta_0) - \frac{2-D\sqrt{2}/v}{\sqrt{D^2-1}}
\left[ \tanh^{-1}\left(\frac{D+\tan\frac{\theta}{2}}{\sqrt{D^2-1}}\right) \right]^\theta_{\theta_0} \nonumber\\
\omega(s-s_0) &=& \frac{\sqrt{1-v^2}}{v}\frac{D\sqrt{2}}{\sqrt{D^2-1}}
\left[ \tanh^{-1}\left(\frac{D+\tan\frac{\theta}{2}}{\sqrt{D^2-1}}\right) \right]^\theta_{\theta_0}
\end{eqnarray}

\subsection{Visualizations} \label{sec:kundtVisuals}

\begin{figure}
\includegraphics[width=120mm]{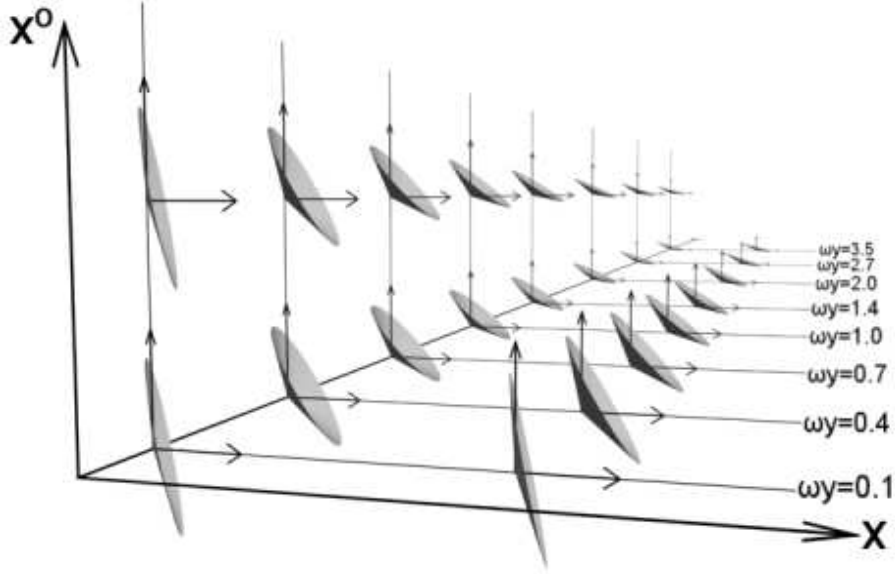}
\caption{A map of the future lightcones of the G\"odel universe in the Kundt coordinates. All the light cones contain the $x$ and $x^0$ axes. The only variation is along $y$: As we go further out in $y$, the cones get stretched along the $x^0$ axis by a factor proportional to $y^{-1}$. }
\label{fig:kundtLightConeDiagram}
\end{figure}

We now capture the above information in a series of diagrams, as we did for the cylindrical system. In Figure \ref{fig:kundtLightConeDiagram}, we display Kundt's version of this space-time's light-cone structure. 

\begin{figure}
\includegraphics[width=120mm]{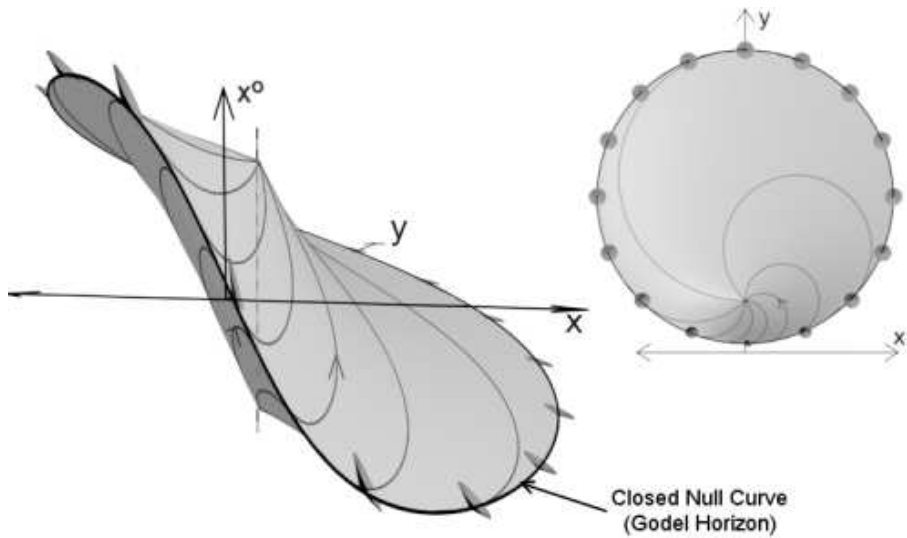}
\caption{The Hawking-Ellis picture in Kundt coordinates: Envelopes of the null geodesics starting from the origin at the same $t$ in different initial directions $\phi_0$. The G\"odel horizon forms a Closed Null Curve. The diagram should be studied in conjunction with the Figures \ref{fig:cylNullGeodesicFamily} and \ref{fig:kundtGeodesics}. The underlying rotational symmetry reveals itself in the refocussing of the photons. The apex of the cone of divergence and that of reconvergence have the same value of $x$ and $y$ (See dotted line). The top view can perhaps be better understood in terms of the $x$-$y$ projections given in Figure \ref{fig:kundtGeodesics}.}
\label{fig:kundtNullGeodesicFamily}
\end{figure}

The translational symmetry along the $x^0$ and $x$ directions is manifest, whereas the cones for different values of $y$ are simply scaled versions of one another. The price we seemingly pay for maintaining manifest homogeneity is that rotationally symmetry all but disappears from the diagram. A vestige of the underlying rotation is nevertheless visible in the definite orientation of all the light-cones, which confers on all geodesics a definite sense (see Figure \ref{fig:kundtGeodesics}). Moreover, even though geodesics diverging from a given point with the same value of $v$ end up describing circles of different radii, they continue to refocus \footnote{The analog of Figure \ref{fig:cylCNCEnvelope} is rather convoluted while that of Figure \ref{fig:cylTimePhiCurves} is very strongly $\phi_0$-dependent. Since neither is very enlightening, we omit them both.} (see Figure \ref{fig:kundtNullGeodesicFamily}).

\begin{figure}
\includegraphics[width=120mm]{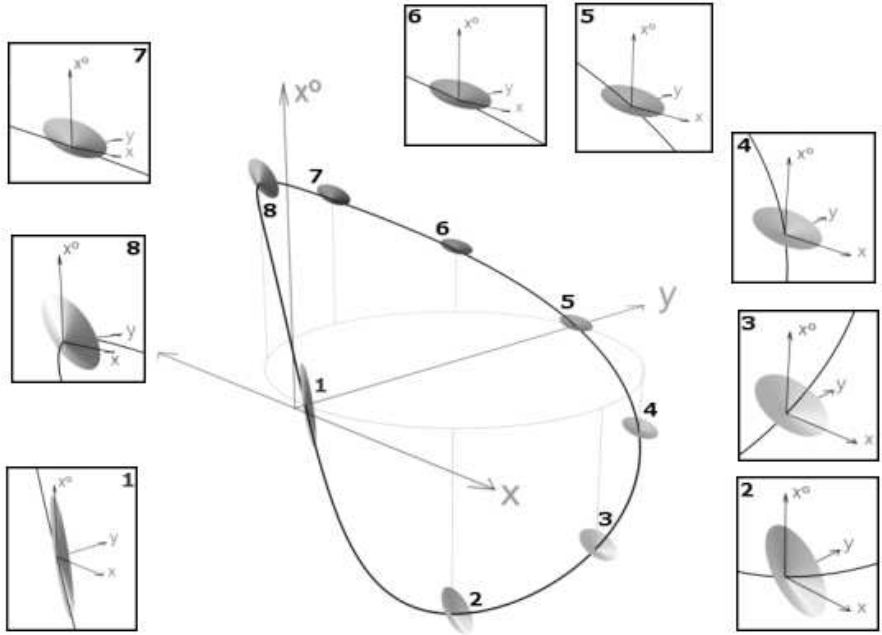}
\caption{The tangent of the Closed Null Curve rotates on the light cone. This diagram should be compared with Figure \ref{fig:cylDetailedCNC} which depicts the generic CNC in the cylindrical system.}
\label{fig:kundtDetailedCNC}
\end{figure}

Note further that {\it all} the future light-cones of this coordinate system contain both the $x^0$- and $x$-axes. We can therefore move back in $x^0$ without moving back locally in time {\it per se},  because whenever we do so, we simultaneously move forward in $x$, which is also time-like. In what ratio we advance in one and retreat in the other depends on where in $y$ we happen to be. By adjusting $D$, we can make the variation in $y$ (and thereby the relative tilt in the light-cones) sufficient over the orbit to allow a CNC/CTC to close. We capture this closure for this system in Figure \ref{fig:kundtDetailedCNC}.

\section{Summary} \label{sec:summary}

To summarize, we have studied worldlines corresponding to motion at constant speed and magnitude of acceleration (WSAs). Through these, we can access any point in the past or future of a comoving observer at the `base-station' where, because of the rotational symmetry of the space-time, these worldlines refocus periodically. We have extended the Hawking \& Ellis depiction of this refocussing to the case of photons subject to a constant acceleration orthogonal to their instantaneous velocities. By adjusting the magnitude of this acceleration we can make the photons refocus at the same point in \emph{both} space and time. 

We have, additionally, studied CNCs/CTCs more general than the ones which develop at/beyond the G\"{o}del horizon. In particular, we have shown how a critical amount of relative tilt in the light-cones of the space-time allows these curves to close, whenever the light-cones in question enclose a spatial axis in addition to the usual temporal one. The critical amount of tilt required for closure detemines the minimal size of the CNC. We have explicitly illustrated this for CNCs of both the cylindrical and the Kundt coordinate systems. Since WSAs lie on cylinders parallel to the time-axis in both systems, we have additionally traced the emergence of CTCs on more easily visualized 2-dimensional plots, by simply cutting open the cylinders, flattening them out and examining the curves they carry as functions of the acceleration. 

Thirdly, we have explored WSAs in their full generality by letting the acceleration range from $-\infty$ to $+\infty$. In particular we have additionally examined both open trajectories, and trajectories that curve with the opposite sense to that of geodesics.

Next, we illustrated all these effects in manifestly homogeneous representation of the G\"{o}del universe in Kundt's coordinates.

Lastly, we have derived formulae relating events in G\"{o}delian space and time to the speed and acceleration of point masses/photons which we would use (in a controlled manner) to travel between them. By carrying out the entire discussion purely in terms of parameters with a direct physical interpretation, we have, we believe, made intuitive several of the otherwise unintuitive properties of the G\"{o}del space-time.

\end{document}